\documentclass[runningheads]{llncs}

\usepackage[T1]{fontenc}
\usepackage{graphicx,verbatim}
\usepackage{booktabs}
\usepackage{multirow}
\usepackage{amsmath}
\usepackage{amssymb}
\usepackage{censor}
\usepackage{hyperref}
\usepackage{enumitem}
\usepackage{xcolor}

\begin{document}

\title{TG-OT: Topology-guided CCTA-IVUS registration via optimal transport matching}
\titlerunning{Topology-guided CCTA-IVUS registration}
%

\author{Rudolf L. M. van Herten\inst{1,2,3} \and
Jos\'e P. Henriques\inst{4} \and
R. Nils Planken\inst{5} \and
Joost~Daemen\inst{6} \and
Eline M. J. Hartman\inst{6} \and
Jolanda J. Wentzel\inst{6} \and
Johannes~C.~Paetzold\inst{1,2} \and
Ivana I\v{s}gum\inst{3,4,5}
}

%
\authorrunning{R. L. M. van Herten et al.}
\institute{Department of Radiology, Weill Cornell Medicine, New York, USA\\\email{rlv4001@med.cornell.edu}\\ \and
Cornell Tech, New York, USA \and
Department of Biomedical Engineering and Physics, Amsterdam UMC, University~of~Amsterdam, Amsterdam, The Netherlands \and
Amsterdam Cardiovascular Sciences, Amsterdam, The Netherlands \and
Department of Radiology, Mayo Clinic, Rochester, USA \and
Department of Cardiology, Erasmus University Medical Center, Rotterdam, The~Netherlands}
  
\maketitle              
\begin{abstract}
Registering coronary CT angiography (CCTA) and intravascular ultrasound (IVUS) enables comprehensive coronary analysis that neither modality can provide alone, yet their fusion remains challenging due to differences in imaging geometry, resolution, and artifact profiles. Existing methods depend on pre-computed lumen or vessel wall segmentations that are unreliable under IVUS acoustic shadowing from calcifications, limiting their clinical applicability. We propose TG-OT, a fully automatic CCTA-IVUS registration framework that eliminates this dependency by integrating trained feature detectors directly into the registration pipeline. Lightweight CNNs are trained to predict calcifications, bifurcations, and lumen radii on the topological $(\theta, z)$ cylinder, encouraging topologically coherent detections without requiring explicit segmentation. Registration is formulated as an optimization over centerline warping parameters, driven by an unbalanced Sinkhorn optimal transport loss on the cylindrical geometry that provides spatially informative gradients even for spatially disjoint predictions, complemented by a lumen matching term. Evaluated on $N{=}47$ paired CCTA-IVUS cases in a 5-fold cross-validation setup, TG-OT achieves strong longitudinal ($\overline{\text{Dice}}_\text{ctl}{=}0.99$), rotational ($\overline{S}_c{=}0.96$), and lumen alignment ($\overline{\text{Dice}}_\text{L}{=}0.69$) without manual interaction or prior segmentation, marking a meaningful step toward clinical integration of automatic CCTA-IVUS fusion.


\end{abstract}

\section{Introduction}

Coronary CT angiography (CCTA) and intravascular ultrasound (IVUS) present complementary modalities for the analysis of coronary artery disease~\cite{komilovich2023coronary,van2024role}: CCTA captures the full 3D arterial geometry non-invasively, while IVUS offers high-resolution cross-sectional imaging of vessel wall morphology and plaque composition~\cite{cury2022cad,mintz2017intravascular}. Registering these modalities enables comprehensive coronary analyses that neither modality can provide alone~\cite{hartman2021lipid,maurovich2022ct}, yet multimodality fusion remains challenging due to fundamental differences in imaging geometry, resolution, and artifact profiles~\cite{haddou2026cyclephase,van2026georeg}.

CCTA-IVUS registration requires extracting a coronary artery centerline from CCTA to generate multiplanar reformatted (MPR) images, which are then aligned to the IVUS pullback. Early methods relied on manual landmark selection and interactive rotation~\cite{van20103d,marquering2008coronary}, while later semi-automatic approaches reduced user burden through seedpoint selection and segmentation~\cite{athanasiou2016three,qian2011intermodal}. More recently, Kadry et al.~\cite{kadry2024morphology} and Li et al.~\cite{li2025autofox} proposed differentiable and Transformer-based frameworks respectively, but both depend on pre-computed lumen or vessel wall segmentations---a condition particularly difficult to satisfy for IVUS, where acoustic shadowing from calcifications obscures vessel wall boundaries~\cite{arora2023state}.

In this work, we propose \textbf{topology-guided image registration via optimal transport (TG-OT)}, a fully automatic CCTA-IVUS registration framework that integrates feature detection directly into the registration pipeline, eliminating the need for prior segmentation. Our contributions are:
\begin{enumerate}[label=(\roman*),nosep,leftmargin=*]
  \item We train lightweight CNNs to predict calcifications, bifurcations, and lumen radii on the topological $(\theta, z)$ cylinder of the artery, circumventing the ill-defined delineation of calcified regions under IVUS acoustic shadowing.
  \item We introduce an unbalanced Sinkhorn optimal transport (OT) loss~\cite{sejourne2023unbalanced} on cylindrical geometry for cross-modal feature matching, stabilizing voxel-wise losses with spatially aware transport costs that tolerate differences in predicted feature mass between modalities.
  \item We demonstrate that the integrated pipeline achieves competitive registration accuracy on $N{=}47$ paired CCTA-IVUS cases without requiring any prior segmentation.
\end{enumerate}

\begin{figure}[t]
    \centering
    \includegraphics[trim={0cm, 0cm, 0cm, 0cm}, clip, width=\columnwidth]{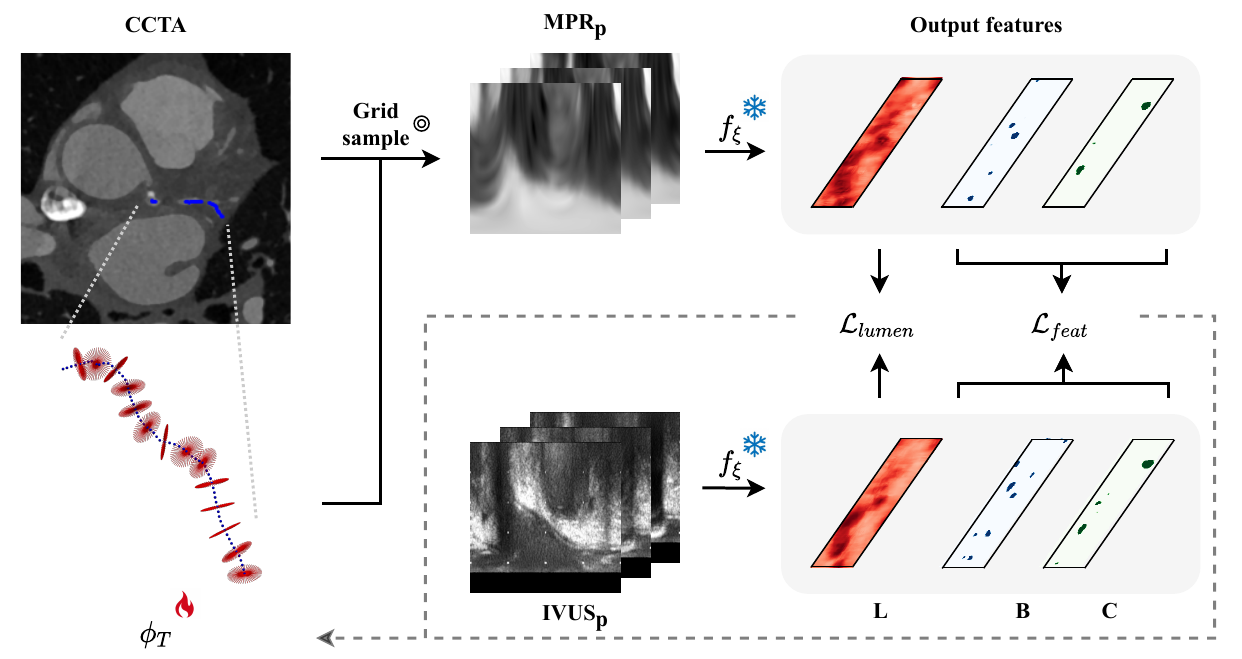}
    \caption{Overview of our TG-OT framework for CCTA-IVUS image registration. A CCTA volume is sampled in polar coordinates along an extracted centerline, with warping governed by trainable parameters $\phi_T$. Frozen CNNs $f_\xi$ predict lumen radii~(L), bifurcations~(B), and calcifications~(C) on the topological cylinder for both modalities, bypassing explicit segmentation. Registration is driven by $\mathcal{L}_\mathrm{feat}$, combining unbalanced Sinkhorn optimal transport with a Dice loss to match sparse features across modalities with spatially informed gradients, and $\mathcal{L}_\mathrm{lumen}$ denoting a lumen radii matching term.}
    \label{fig:overview}
\end{figure}

\section{Method}

We present TG-OT, a flexible framework for automatic CCTA-IVUS registration. Our method comprises two main components. First, CNNs are trained to identify key anatomical features along with lumen radii from polar-transformed MPR or IVUS data, while additionally learning to detect guidewire artifacts in IVUS images. Second, once trained, the CNNs are integrated into a differentiable image registration module which optimizes a set of transformation parameters that optimally align an IVUS pullback to a corresponding automatically extracted coronary artery centerline in CCTA~\cite{hampe2024graph}. An overview of our registration method is presented in Fig.~\ref{fig:overview}. Code is available on~\href{https://github.com/RoelvH97/TG_OT}{\color{blue}GitHub}.

\subsection{Feature detection CNN}

\begin{figure}[t]
    \centering
    \includegraphics[trim={0cm, 0cm, 0cm, 0cm}, clip, width=\columnwidth]{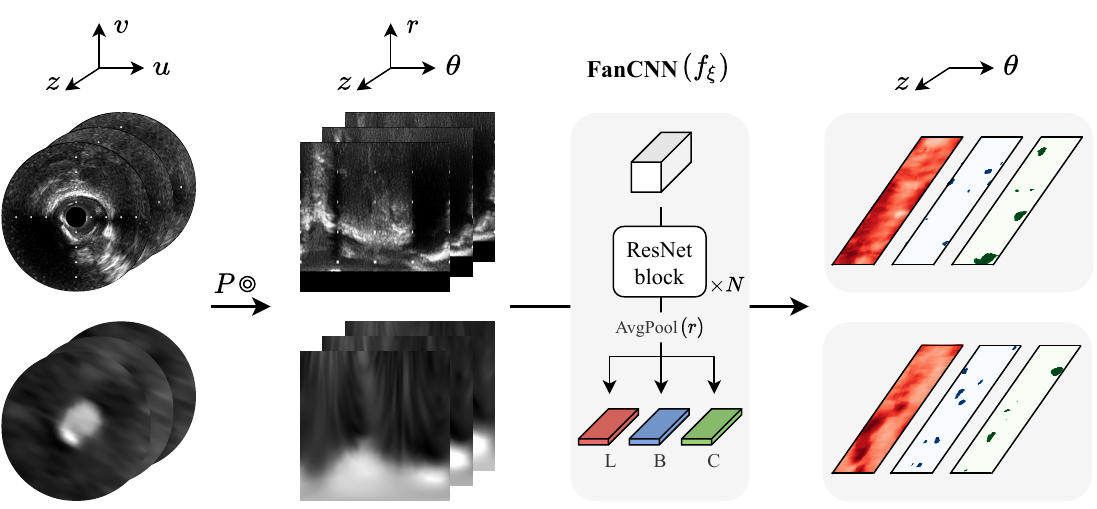}
    \caption{Overview of the feature detection CNN $f_{\xi}$. IVUS (top) and MPR (bottom) data are transformed from Cartesian ($u$, $v$, $z$) to polar ($r$, $\theta$, $z$) coordinates via $P~\odot$. A 3D CNN consisting of $N$ ResNet blocks processes the polar-transformed data, followed by average pooling along the radial ($r$) dimension to produce 2D ($\theta$, $z$) predictions of lumen radii (L), bifurcation locations (B), and calcification presence (C). The resulting $(\theta, z)$ map represents the unwound topological cylinder $\mathbb{S}^1 \times \mathbb{R}$, with periodic boundary conditions at $\theta = 0 \equiv 2\pi$.}
    \label{fig:polar_classifier}
\end{figure}

Aligning CCTA and IVUS requires identifying anatomical features visible across both modalities. We detect lumen radii, calcifications, and bifurcation locations from polar-transformed MPR and IVUS data, operating in the polar $(r,\theta,z)$ domain where classifying the \emph{presence} of structures along each radial ray is well-defined, even when full segmentation is ill-posed under IVUS acoustic shadowing. We adapt FanCNN~\cite{van2023automatic} and train separate networks $f_\xi$ for MPR and IVUS, with the IVUS network additionally predicting guidewire artifacts. Each $f_\xi$ is supervised with a combination of soft Dice and Betti matching loss~\cite{berger2024topologically,stucki2023topologically} ($\alpha_\text{BM} = 0.05$), the latter penalizing topological errors in detected features, a useful property for reliable downstream OT matching. Lumen radii are simultaneously regressed using a standard $L_1$ loss. An overview is shown in Fig.~\ref{fig:polar_classifier}.

\subsection{Registration pipeline}
\label{sec:registration}
Registration is formulated as a centerline warping problem: given a CCTA centerline and an IVUS pullback, we optimize transformation parameters that align the two through a combination of sparse anatomical feature matching, intensity-based similarity, and lumen area agreement. The pipeline proceeds in three stages: a transformation model defines the warping parameterization, rigid pre-alignment initializes global pose, and deformable optimization jointly refines all parameters, driven primarily by an unbalanced OT objective on the cylindrical geometry. We denote the CCTA-derived MPR as the warped moving image $M_w$ and the IVUS pullback as the fixed image $F$ throughout.

\subsubsection{Transformation model.}

Given a CCTA centerline $\mathcal{C} \in \mathbb{R}^{n \times 3}$ and an IVUS pullback, we seek transformation parameters $\phi_T$ comprising global scaling $s_z$ and translation $t_z$ along the vessel axis, and per-frame local rotation $\theta_p$ and in-plane displacement $(t_{u,p}, t_{v,p})$. Following~\cite{kadry2024morphology}, local rotations are parameterized via a cumulative B-spline ~\cite{eilers1996flexible}, naturally enforcing smoothness and capturing the gradual twist of the IVUS catheter during pullback. Each centerline point $\mathbf{x}_p \in \mathbb{R}^3$ is transformed by:
\begin{equation}
    \mathbf{x}'_p = \mathbf{T}_g \, \mathbf{T}_{l,p} \, \mathbf{x}_p,
\end{equation}
where $\mathbf{T}_g \in \mathbb{R}^{4\times4}$ embeds the global 1D affine parameters $(s_z, t_z)$ into a 3D transformation, and $\mathbf{T}_{l,p} \in \mathrm{SE}(2)$ encodes local in-plane rotation and displacement, embedded analogously with an identity component along $z$.

\subsubsection{Rigid pre-alignment.}
Pre-alignment is performed in two stages. First, the global scaling factor $s_z$ is initialized analytically as the ratio of the CCTA centerline length to the IVUS pullback length. Longitudinal alignment is then determined by grid search over $t_z$ while minimizing a combined objective:
\begin{equation}
    \mathcal{L}_\mathrm{init} = \sum_{k \in \{B,C\}} \mathcal{L}_\mathrm{Dice}\bigl(\ell^k_{M_{w}}(z),\, \ell^k_F(z)\bigr) + \mathcal{L}_\mathrm{lumen}(A_{M_{w}}, A_F),
\end{equation}
where $\ell^k(z) = \max_\theta f_\xi\bigl(P(\cdot)\bigr)_k(\theta, z)$ aggregates the 2D classifier output for class $k$ along the angular dimension to produce a 1D longitudinal signal per feature type, $\mathcal{L}_\mathrm{Dice}$ denotes the soft Dice loss, and $\mathcal{L}_\mathrm{lumen}$ matches cross-sectional lumen areas~($A$) computed from predicted radii via a radial polygon approximation, using a combination of absolute error and normalized cross-correlation~\cite{avants2008symmetric}. Subsequently, $\theta_0 \in [0, 2\pi)$ is found by grid search (step size $=5^\circ$), shifting classifier outputs along $\theta$ and evaluating $\mathcal{L}_\text{Dice}$ summed over $k \in \{B,C\}$.

\subsubsection{Deformable optimization.}

Starting from the rigid initialization, all parameters $\phi_T$ are jointly refined. The objective combines sparse anatomical feature matching, lumen radius agreement, and parameter regularization:

\begin{equation}\label{eq:total_loss}
    \hat{\phi}_T = \arg\min_{\phi_T} \bigl[\mathcal{L}_\mathrm{feat} + \alpha_\mathrm{lumen}\,\mathcal{L}_\mathrm{lumen} + \alpha_\mathrm{reg}\,\mathcal{L}_\mathrm{reg}\bigr].
\end{equation}
where $\mathcal{L}_\text{feat}$ is the feature matching term detailed below, $\mathcal{L}_\text{lumen}$ provides complementary lumen-based alignment, and $\mathcal{L}_\text{reg}$ enforces spatial smoothness of the local transformation parameters.

\subsubsection{Feature matching via unbalanced optimal transport.}

A voxel-wise loss such as Dice penalizes feature mismatch uniformly regardless of spatial proximity, providing no gradient signal when predicted features do not overlap. We instead formulate feature matching as an optimal transport problem on the topological cylinder $\mathbb{S}^1 \times \mathbb{R}$, where the transport cost encodes geodesic distance and thus provides informative gradients even for spatially disjoint predictions.

Given classifier probability maps derived from polar-transformed $P(\cdot)$ data
\begin{equation}\label{eq:classifiers}
    \mu^k = f_\xi\bigl(P(M_w)\bigr)_k, \qquad \nu^k = f_\xi\bigl(P(F)\bigr)_k, \qquad k \in \{B, C\},
\end{equation}
for the warped moving image $M_w$ and fixed image $F$ respectively, the feature matching loss combines an OT term with a Dice overlap term:
\begin{equation}\label{eq:feat_loss}
    \mathcal{L}_\mathrm{feat} = \sum_{k \in \{B,C\}} \bigl[\mathcal{L}_\mathrm{Dice}(\mu^k, \nu^k) + \alpha_\mathrm{OT}\,\mathcal{L}_\mathrm{OT}(\mu^k, \nu^k)\bigr].
\end{equation}
The Dice term encourages overlap between predicted feature mass per class, while the OT term captures the spatial cost of transporting misaligned features across the cylinder.

\paragraph{Transport cost.}
The cost matrix is defined via geodesic cylindrical distance:
\begin{equation}
    C_{ij} = \sqrt{d_\theta(i,j)^2 + d_z(i,j)^2},
\end{equation}
where $d_\theta$ is the circular (wraparound) distance on $\mathbb{S}^1$, and $d_z$ is the Euclidean distance along the vessel axis.

\paragraph{Unbalanced Sinkhorn formulation.}
The optimal transport plan is computed via the Sinkhorn algorithm~\cite{cuturi2013sinkhorn} with entropic regularization $\varepsilon$ and unbalanced marginal relaxation parameter $\tau \in (0, 1)$:
\begin{equation}
    \mathcal{L}_\mathrm{OT} = \langle \mathbf{C},\, \mathbf{P}^* \rangle, \qquad \text{where} \quad \mathbf{P}^* = \operatorname{diag}(\mathbf{u})\, \mathbf{K}\, \operatorname{diag}(\mathbf{v}),
\end{equation}
with Gibbs kernel $K_{ij} = \exp(-C_{ij}/\varepsilon)$, marginals $\boldsymbol{\mu}, \boldsymbol{\nu}$ given by the per-class feature maps $\mu^k, \nu^k$ of Eq.~(\ref{eq:classifiers}), and scaling vectors $(\mathbf{u}, \mathbf{v})$ obtained by iterating:
\begin{equation}
    \mathbf{u} \leftarrow \left(\frac{\boldsymbol{\mu}}{\mathbf{K}\mathbf{v}}\right)^{\!\tau}, \qquad
    \mathbf{v} \leftarrow \left(\frac{\boldsymbol{\nu}}{\mathbf{K}^\top\mathbf{u}}\right)^{\!\tau}.
\end{equation}
Here $\tau \in (0,1)$ directly controls marginal relaxation; the corresponding KL divergence penalty weight on the marginal constraints is $\rho = \varepsilon\tau/(1 - \tau)$~\cite{sejourne2023unbalanced}, with $\tau \to 1$ recovering balanced OT. We utilize the unbalanced formulation, as the CNN may predict features more prominently in one modality than the other due to image resolution and contrast discrepancies, so enforcing exact marginal constraints would penalize legitimate asymmetries in detection confidence.

\subsubsection{Auxiliary losses.} Lumen radius agreement is measured via normalized cross-correlation $\mathcal{L}_\text{NCC}$ between the predicted radii of the warped MPR and IVUS, providing an area-based alignment signal in pullback segments with sparse anatomical landmarks. Both $\mathcal{L}_\text{NCC}$ and image similarity terms are computed exclusively over regions without guidewire interference, as identified by the IVUS guidewire predictions of $f_\xi$. Additionally, the regularization term $\mathcal{L}_\text{reg}$ penalizes the squared gradient of local parameters $\{\theta_p, t_{u,p}, t_{v,p}\}$ to enforce spatial smoothness.

\begin{table}[t]
\centering
\caption{Cylindrical feature detection performance of $f_{\xi}$ on IVUS and MPR across training loss configurations. Reported metrics are Dice score for lumen, bifurcation, calcification, and guidewire detection, Betti matching error ($\tau_{\text{err}}$) for bifurcations and calcifications, and centerline Dice (clDice) for guidewire continuity. Results are reported as mean~(std) over 5-fold cross-validation, with the best score indicated in \textbf{bold}.}
\fontsize{8pt}{11pt}\selectfont
\setlength{\tabcolsep}{0.35pt}  
\begin{tabular}{llccccccc}
\toprule
   & & Lumen & \multicolumn{2}{c}{Bifurcations} & \multicolumn{2}{c}{Calcifications} & \multicolumn{2}{c}{Guidewire} \\
   \cmidrule(lr){3-3} \cmidrule(lr){4-5} \cmidrule(lr){6-7} \cmidrule(lr){8-9}
   & Loss & Dice $\uparrow$ & Dice $\uparrow$ & $\tau_{\mathrm{err}} \downarrow$ & Dice $\uparrow$ & $\tau_{\mathrm{err}} \downarrow$ & Dice $\uparrow$ & clDice $\uparrow$ \\
\midrule
\multirow{4}{*}{\rotatebox{90}{IVUS}} & $\mathcal{L}_{\mathrm{BCE}}$ & \textbf{0.94} (0.02) & 0.46 (0.17) & 4.53 (3.33) & \textbf{0.68} (0.25) & 4.34 (3.79) & 0.55 (0.17) & 0.60 (0.19)\\
 & $\mathcal{L}_{\mathrm{Dice}}$ & 0.90 (0.04) & 0.53 (0.14) & 4.53 (2.38) & 0.64 (0.26) & \textbf{3.66} (3.60) & 0.67 (0.08) & 0.71 (0.11)\\
 & $\mathcal{L}_{\mathrm{BCE+Dice}}$ & 0.92 (0.02) & 0.52 (0.16) & 4.23 (2.71) & 0.67 (0.24) & 4.09 (3.83) & 0.64 (0.14) & 0.68 (0.16) \\
 & $\mathcal{L}_{\mathrm{BM+Dice}}$ & 0.92 (0.02) & \textbf{0.56} (0.13) & \textbf{4.15} (2.65) & 0.67 (0.23) & 3.89 (2.98) & \textbf{0.69} (0.09) & \textbf{0.72} (0.11) \\
\midrule
\multirow{4}{*}{\rotatebox{90}{MPR}} & $\mathcal{L}_{\mathrm{BCE}}$ & 0.90 (0.04) & 0.53 (0.17) & 2.61 (1.87) & 0.60 (0.27) & 1.65 (2.34) & - & - \\
 & $\mathcal{L}_{\mathrm{Dice}}$ & 0.90 (0.04) & \textbf{0.62} (0.13) & \textbf{2.33} (1.71) & \textbf{0.65} (0.25) & 1.45 (1.87) & - & - \\
 & $\mathcal{L}_{\mathrm{BCE+Dice}}$ & 0.89 (0.05) & 0.59 (0.16) & 2.37 (1.80) & 0.62 (0.26) & 1.65 (2.14) & - & - \\
 & $\mathcal{L}_{\mathrm{BM+Dice}}$ & \textbf{0.91} (0.03) & 0.61 (0.12) & \textbf{2.33} (1.89) & 0.63 (0.26) & \textbf{1.39} (1.94) & - & - \\
\bottomrule
\end{tabular}
\label{tab:cnn_results}
\end{table}

\section{Experiments}

\subsubsection{Data.}
All experiments are conducted on a subset of the IMPACT study $(N{=}47)$ from the Erasmus University Medical Center~\cite{hartman2021lipid} for which both CCTA and IVUS data are available. For all IVUS pullbacks, a reference CCTA coronary artery centerline with tangent orientation was manually identified by matching large side branches visible in both modalities. Lumen segmentations were made available for both modalities, from which lumen radii were derived in the polar domain. Additionally, the presence of calcifications and bifurcations was manually annotated for each radial direction in the polar-transformed data, yielding binary labels on the $(\theta,z)$ cylinder for both MPR and IVUS data.

\subsubsection{Evaluation.}
\label{sec:eval_reg}

All experiments were evaluated in a 5-fold cross-validation setup, ensuring metrics are reported for patients unseen during classifier training. Since no directly comparable fully automatic CCTA-IVUS registration method exists, and direct comparison with semi-automatic approaches such as Kadry et al.~\cite{kadry2024morphology} is precluded by segmentation label unavailability and modality mismatch, we evaluate against two natural baselines: the unregistered case, and a standard Dice feature matching loss in place of the proposed OT objective. We additionally ablate the lumen alignment term and evaluate NMI as an image similarity term $\mathcal{L}_{\text{image}}$~\cite{pluim2003mutual}. CNN training loss configurations ($\mathcal{L}_\text{BCE}$, $\mathcal{L}_\text{Dice}$, $\mathcal{L}_\text{BCE+Dice}$, $\mathcal{L}_\text{BM+Dice}$) are evaluated separately, reporting Dice and Betti matching error $\tau_\text{err}$ in dimension-0~\cite{berger2025pitfalls,stucki2023topologically} for bifurcation and calcification detection, and clDice~\cite{shit2021cldice} for guidewire continuity. Registration accuracy is assessed by centerline overlap ($\text{Dice}_\text{ctl}$) following Schaap et al.~\cite{schaap2009standardized}, cosine similarity of cross-sectional normal vectors ($S_c$), and lumen overlap ($\text{Dice}_\text{L}$), reported as median (Q1, Q3). 

We set $\alpha_{\text{OT}}{=}0.1$, $\alpha_{\text{lumen}}{=}1.0$, $\alpha_{\text{image}}{=}10^ 2$, $\alpha_{\text{reg}}{=}10^ 3$, $\varepsilon{=}0.1$, and $\tau{=}0.8$ for all registration experiments where the respective loss terms are included.

\begin{table}[t]
\centering
\caption{Registration results across longitudinal initialization (\textit{z}$_\text{init}$), rotational initialization ($\theta_\text{init}$), and deformable refinement. Rows marked $^\dagger$ are baselines representing standard registration approaches; remaining rows ablate components of the proposed method. The best prior-stage configuration is carried forward at each stage; best results are denoted in \textbf{bold}. Metrics are centerline overlap Dice$_\text{ctl}$, cosine similarity of cross-sectional normal vectors $S_c$, and lumen overlap Dice$_\text{L}$, reported as median (Q1, Q3) over 5-fold cross-validation. The top row (--) denotes the unregistered baseline.}
\fontsize{8pt}{11pt}\selectfont
\setlength{\tabcolsep}{1pt}  
\begin{tabular}{llllccc}
\toprule
   & $\mathcal{L}_{\mathrm{feat}}$ & $\mathcal{L}_{\mathrm{lumen}}$ & $\mathcal{L}_{\mathrm{image}}$ & Dice$_{\mathrm{ctl}} \uparrow$ & $S_\mathrm{c} \uparrow$ & Dice$_\mathrm{L} \uparrow$\\
\midrule
- & - & - & - & 0.67 (0.56, 0.81) & 0.21 (-0.36, 0.55) & 0.47 (0.40, 0.55) \\
\midrule
\multirow{3}{*}{$z_{\mathrm{init}}$} & Dice$^\dagger$ & - & - & 0.97 (0.87, 1.00) & 0.16 (-0.54, 0.64) & 0.57 (0.49, 0.62) \\
& Dice$^\dagger$ & NCC & - & 0.98 (0.93, 1.00) & 0.16 (-0.56, 0.69) & 0.58 (0.52, 0.62) \\
& Dice & NCC + L$_1$ & - & \textbf{0.99} (0.97, 1.00) & 0.11 (-0.56, 0.65) & 0.58 (0.54, 0.64) \\
\midrule
\multirow{3}{*}{$\theta_{\mathrm{init}}$} & Dice$^\dagger$ & - & - & 0.99 (0.97, 1.00) & 0.94 (0.75, 0.97) & 0.63 (0.56, 0.68) \\
& OT & - & - & 0.99 (0.97, 1.00) & 0.94 (0.84, 0.98) & 0.63 (0.56, 0.68) \\
& OT + Dice & - & - & 0.99 (0.97, 1.00) & \textbf{0.95} (0.87, 0.97) & 0.63 (0.56, 0.68) \\
\midrule
\multirow{5}{*}{deform} & Dice$^\dagger$ & - & NMI & 0.98 (0.96, 1.00) & 0.95 (0.86, 0.97) & 0.66 (0.59, 0.71) \\
& Dice$^\dagger$ & NCC & - & \textbf{0.99} (0.97, 1.00) & \textbf{0.96} (0.90, 0.98) & 0.68 (0.65, 0.74) \\
& Dice$^\dagger$ & NCC & NMI & 0.98 (0.96, 1.00) & 0.95 (0.71, 0.97) & \textbf{0.69} (0.64, 0.74)\\
& OT & NCC & - & \textbf{0.99} (0.97, 1.00) & 0.94 (0.80, 0.97) & 0.68 (0.65, 0.74) \\
& OT + Dice & NCC & - & \textbf{0.99} (0.97, 1.00) & \textbf{0.96} (0.92, 0.98) & \textbf{0.69} (0.64, 0.74) \\
\bottomrule
\end{tabular}
\label{tab:reg_results}
\end{table}

\begin{figure}[t]
    \centering
    \includegraphics[trim={0cm, 0cm, 0.75cm, 1cm}, clip, width=\columnwidth]{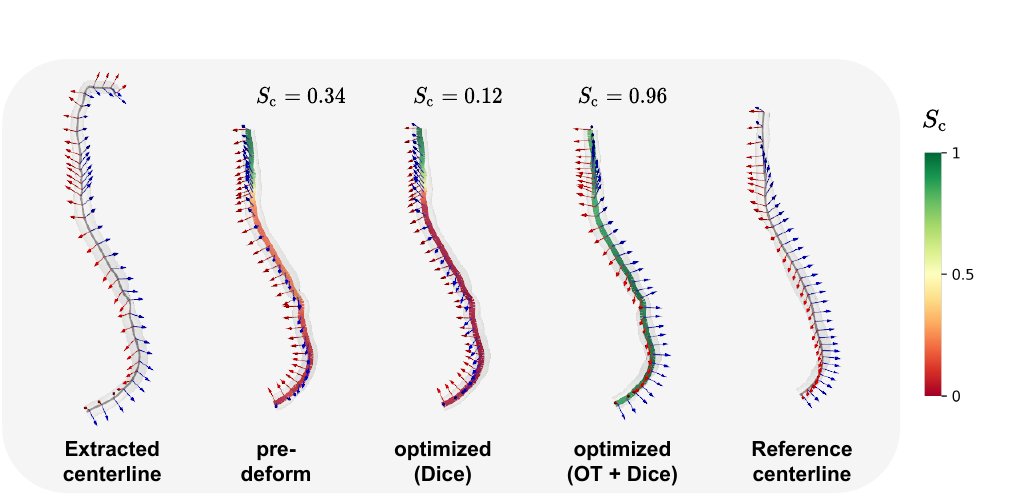}
    \caption{Qualitative registration result for a representative case exhibiting catheter rotational drift. Cross-sectional normal vectors are color-coded by cosine similarity $S_c$. Columns show, from left to right: the extracted CCTA centerline (unregistered), the pre-deformable alignment, deformable optimization with Dice-only feature matching, deformable optimization with OT+Dice feature matching, and the reference centerline. The Dice-only model fails to recover rotational alignment throughout the pullback, as the rigid rotation initialized at $\theta_\text{init}$ captures only part of the catheter twist. The OT+Dice model successfully corrects this drift.}
    \label{fig:example}
\end{figure}

\section{Results}

\subsubsection{Feature detection.}
\label{sec:results_feature}
Table~\ref{tab:cnn_results} compares classification loss configurations. Lumen Dice is stable across configurations, as expected given the shared $\mathcal{L}_1$ regression objective. For bifurcations and calcifications, $\mathcal{L}_{\text{BM+Dice}}$ consistently yields low Betti matching errors across modalities while maintaining competitive Dice scores, indicating topologically coherent detections critical for downstream OT matching~\cite{lux2025topograph}. We therefore adopt $\mathcal{L}_{\text{BM+Dice}}$ for all registration experiments.

\subsubsection{Registration.}
\label{sec:results_registration}

Table~\ref{tab:reg_results} reports registration accuracy across all conditions. Rotational initialization ($\theta_\text{init}$) identifies the approximate global minimum of the rotational objective, placing subsequent deformable optimization in the correct basin of convergence. The Dice-only baseline achieves competitive performance when combined with NCC lumen matching (Dice$_\text{ctl}$: 0.99, $S_c$: 0.96, Dice$_\text{L}$: 0.68), while adding $\mathcal{L}_\text{NMI}$ does not improve results, consistent with limited contrast correspondence between MPR and IVUS. The proposed OT+Dice objective yields a statistically significant improvement in rotational alignment over the Dice-only baseline (Wilcoxon signed-rank $p<0.01$), driven by more consistent performance across the cohort ($S_c$ Q1: $0.92$ vs.\ $0.90$) and successful recovery of cases with severe catheter rotational drift where the Dice-only baseline fails (see Fig.~\ref{fig:example}). Optimization took $90$ seconds per case on a single RTX 2080 Ti GPU.

\section{Discussion and Conclusion}

TG-OT achieves strong CCTA-IVUS registration performance without requiring any manual interaction or prior 3D image segmentation, outperforming Dice-only and NMI baselines in rotational accuracy---particularly for cases with severe catheter drift, where voxelwise losses lack the spatial gradients to recover alignment. Failure cases were observed in two categories: longitudinal initialization failed in $N{=}4$ cases due to incorrectly truncated automatic centerline extraction from CCTA, while rotational alignment degraded in $N{=}2$ cases with extreme landmark sparsity. Furthermore, because registration is driven by features detected on the resampled CCTA volume, its reliability is bounded by that of the feature detection step, for which foundation models may offer greater robustness at scale~\cite{carion2025sam,scheinfeld2026multimodal}. Nevertheless, all $N{=}47$ cases are included in the reported metrics. A broader challenge is the absence of automatic reference standards for CCTA-IVUS registration, making the constructed baselines the most meaningful available comparison. By removing the dependency on error-prone 3D intermediate segmentations, TG-OT addresses a key barrier to scalable multimodal coronary analysis and brings fully automatic CCTA-IVUS fusion closer to clinical deployment.

\subsubsection{Disclosure of Interests.} Ivana I\v{s}gum has received research funding from Philips Healthcare, the Dutch Research Council, Pie Medical Imaging BV, Esaote Europe BV, Abbott Vascular Inc, Horizon Europe, and Innovative Health Initiative. Jos\'{e} P. Henriques has received research funding from Health Holland, Infraredx Inc, B. Braun Medical Inc, and Nipro Medical Europe NV. All other authors declare no competing interests.

\bibliographystyle{splncs04}
\bibliography{Paper-1931}

\end{document}